\documentclass{Interspeech2024}

\interspeechcameraready

\title{Feature Representations for Automatic Meerkat Vocalization Classification}

\name[affiliation={1}]{Imen}{Ben Mahmoud}
\name[affiliation={1,2}]{Eklavya}{Sarkar}
\name[affiliation={3}]{Marta}{Manser}
\name[affiliation={1}]{Mathew}{Magimai.-Doss}

\address{
  $^1$Idiap Research Institute, Martigny, Switzerland\\
  $^2$École polytechnique fédérale de Lausanne (EPFL), Switzerland \\
  $^3$University of Zurich (UZH), Switzerland}
\email{\{ibmahmoud, esarkar, mathew\}@idiap.ch, marta.manser@ieu.uzh.ch}


\begin{document}

\maketitle

\begin{abstract}
Understanding evolution of vocal communication in social animals is an important research problem. In that context, beyond humans, there is an interest in analyzing vocalizations of other social animals such as, meerkats, marmosets, apes. 
While existing approaches address vocalizations of certain species, a reliable method tailored for meerkat calls is lacking. To that extent, this paper investigates feature representations for automatic meerkat vocalization analysis. Both traditional signal processing-based representations and data-driven representations facilitated by advances in deep learning are explored. Call type classification studies conducted on two data sets reveal that feature extraction methods developed for human speech processing can be effectively employed for automatic meerkat call analysis. 
\end{abstract}

\section{Introduction}
Meerkats are highly social animals with a complex social structure~\cite{social-meerkats}. Featuring a dominant breeding pair and cooperative behaviors, they dig safe places through their foraging areas. Communication among a clan occurs through various vocalizations including barks, chirps, trills, and growls. They are essential in coordinating group activities, warning of potential dangers, and maintaining social cohesion. Researchers have identified and classified around 30 types of vocalizations in meerkats~\cite{townsend2014}. These vocalizations can be categorized into alarm calls emitted when a potential predator is encountered~\cite{alarm-calls}, contact calls used to maintain group cohesion~\cite{close-calls}, and dominance calls employed during a conflict to assert social hierarchy. Additional vocalizations serve to express various other emotions. These vocalizations are part of a complex communication system, influenced by the group's social organization and ecology~\cite{MANSER2014}.

Over the past two decades, there has been a notable improvement in understanding this communication system, particularly in decoding the context of calls. For example, in~\cite{MANSER2002}, it is demonstrated that meerkat alarm calls encode information about both predator type and the signaler’s perception of urgency simultaneously. Additionally, in~\cite{gall2017group}, it was found that close calls are used to adjust movement direction and maintain group cohesion, especially in low-visibility environments and during continuous movement. However, understanding the context precedes contextual analysis. The process of categorizing calls is mainly conducted by human listeners, who rely on their expertise. Nonetheless, even among these experts, varying interpretations may arise, highlighting the complexity inherent in the classification task~\cite{Kershenbaum}.

Although previous research has provided insights into the social and contextual aspects of meerkat vocalizations, there remains a lack of computational methods for the automatic analysis of this language. Specifically, to the best of our knowledge, there has not been a formal study on the automatic classification of meerkat vocalizations. One of the main reasons being that biological level and linguistic level analysis of meerkat vocalizations has evolved more recently, leading to the availability of reliable data sets for automatic analysis. As a first step, the present paper aims to investigate feature representations for automatic meerkat vocalization analysis. The motivation for this arises from the important role feature representation plays in pattern analysis and classification systems. In the past, in the field of speech and audio processing, these representations were largely obtained by combining prior knowledge with signal processing. Even though meerkat vocalizations have been analyzed using signal processing, there is still a lack of reliable prior knowledge to extract feature representations for automatic analysis. In recent years, with advances in deep learning, data-driven feature representations have become more prominent and have been demonstrated useful for bioacoustic analysis. In this paper, we investigate both types of feature representations. 


The remainder of the paper is organized as follows: Section~\ref{sec:feature-representation} introduces the two types of feature representations, providing a detailed overview of the methods used. Section~\ref{sec:experiments} delineates the experimental setup and workflow, including the dataset used during the study, the classification setup, and the evaluation metric. Section~\ref{sec:results-discussion} presents the classification results with a comprehensive analysis of the findings. Finally, Section~\ref{sec-conclusions} concludes our study.

\section{Feature representations}\label{sec:feature-representation}

This section motivates and presents the different feature representations investigated in this paper. These representations are grouped as (a) knowledge-based/hand-crafted feature representations and (b) neural-based data-driven feature representations.

\subsection{Knowledge-based/hand-crafted feature representations}

\textbf{Catch22}: Highly Comparable Time-Series Analysis (HCTSA) is an interpretable signal processing-based framework, where a set of 7700 features are extracted by characterizing the signal by different time series analysis methods, such as linear correlation, modeling fitting (e.g., autoregressive moving average analysis, GARCH), wavelet analysis, and extraction of information theoretic measures. It is then combined with feature selection to build statistical models for the end task~\cite{hcsta}. The efficacy of this framework has been demonstrated for bioacoustic analysis. For instance, these features have been investigated for behavioral birdsong discrimination~\cite{hcsta_birdsong}, automated acoustic monitoring of ecosystems~\cite{sethi2020automated}, as well as marmoset caller identification~\cite{hcsta_marmosat}. One of the limitations of the HCTSA approach is computational complexity, as it involves the evaluation of many similar features. In recent work, CAnonical Time-series CHaracteristics (Catch22) features, a subset of 22 HCTSA features that are minimally redundant has been proposed, and its utility has been demonstrated across 93 real-world time-series classification problems~\cite{catch22}. 
These features fall into different conceptual grouping such as distribution shape, linear autocorrelation, incremental differences, and self-affine scaling. The dimension of the feature set is 24 including the mean and the standard deviation. 

\noindent
\textbf{COMPARE}: COMPARE features have been developed for paralinguistic speech processing. The COMPARE feature set of length 6373 consist of functionals of (a) energy related low level descriptors (LLDs), (b) spectral LLDs, and (c) voicing related LLDs estimated over an utterance~\cite{compare}. 

\noindent
\textbf{eGeMAPS}: extended Geneva Minimalistic Acoustic Parameter
Set (eGeMAPS) is yet another feature set developed for paralinguistic speech processing~\cite{egemaps}. The feature set consists of 88 different features. They are obtained by extracting (a) LLDs, namely, frequency-related parameters, energy/amplitude related parameters, and spectral (balance) parameters, and (b) temporal features consisting of the rate of loudness peaks, mean length and standard deviation of voiced and unvoiced regions, and number of continuous voices regions per second from the acoustic signal.

\subsection{Neural-based data-driven feature representations}

\textbf{Self-supervised learning-based}: In traditional supervised learning, models rely on labeled data, which is expensive and time-consuming to obtain. Thus, the emergence of self-supervised learning (SSL) techniques offers a powerful alternative to these learning methods by leveraging unlabeled data and designing pretext tasks involving human speech. By doing so, it allows models to learn meaningful representations without relying on explicit human annotations. In~\cite{interspeech-eklavya}, the authors explored leveraging embedding spaces focusing on the Marmoset caller discrimination. The study demonstrated that representations pre-trained on human speech could be effectively applied to the bio-acoustics domain. Motivated by that study, we chose three popular SSL models, namely, WavLM~\cite{wavlm}, wav2vec2~\cite{wav2vec} and HuBERT~\cite{hubert}, pre-trained with 960 hours of audio from Librispeech corpus~\cite{librispeech}. We extract embeddings from one of the layers or all layers of the SSL model and model it for call classification.


\noindent
\textbf{Supervised-learning based} (denoted as CNN-crafted): In this part, we focus on the feature extraction phase within a classification framework. This involves directly inputting waveform data into a neural network using an end-to-end Convolutional Neural Network (CNN) architecture.
The architecture is inspired by~\cite{palaz2019} and is presented in Table~\ref{tab:cnn_architecture}. The model is trained to perform call type classification. After training, we derive a feature set of dimension 80 from each call by extracting the output of the penultimate layer of the model, referred to as CNN handcrafted features throughout the study.

\section{Experiments}\label{sec:experiments}

This section presents the dataset of our study, consisting of two Sets ( A and B ) of meerkat calls used during the study, followed by a detailed breakdown of the study's workflow.

\subsection{Meerkat calls dataset}
Set A consists of 90 audio recordings of 9 different meerkat call types collected and labeled by Prof. Marta Manser, University of Zurich, following ethical approval: Aggression (agg), Sentinel (sen), Alarm (al), Chatter (ch), Grooming (gr), Close-call (cc), Submission (sub), Lead (ld) and Sunning (su). Every file was manually segmented using Koe~\cite{koe}; an open-source software to visualize, segment, and classify acoustic units in animal vocalizations, amounting to a total of 1795 vocalization segments at a sampling rate of \SI{44.1}{\kilo\hertz}, with a mean and median length of \SI{161(118)}{\milli\second} and \SI{102}{\milli\second} respectively. Table~\ref{tab:setAdistribution} shows the distribution of the different call types of Set A. It is crucial to emphasize that this table reveals a significant imbalance within the dataset, mirroring the real-world scenario.
\begin{table}[!htb]
    \centering
    \caption{Distribution of the different call types present in Set A.}
    \label{tab:setAdistribution}
    \begin{tabular}{|ccccccccc|}
    \hline
         \textbf{agg} & \textbf{sen} & \textbf{al} & \textbf{ch} & \textbf{gr} & \textbf{cc}  & \textbf{sub}  & \textbf{ld} & \textbf{su} \\
         \hline
         125& 411  & 609 & 108 & 12  & 81 & 99 & 28 & 322\\
         \hline
    \end{tabular}
   
\end{table}

Set B is a public dataset~\cite{mara-data}. The corpus consists of 6428 individual files, categorized into 7 call types, sampled at \SI{48}{\kilo\hertz} with a mean of \SI{148(96)}{\milli\second} and a median of \SI{124}{\milli\second}. Four classes seen previously in Set A are also present in Set B, with three additional ones: Short note (sn), Social call (sc), and Move (mv). Table~\ref{tab:setBdistribution} displays the distribution of the different call types in Set B.

\begin{table}[!htb]
 \centering

 \caption{Distribution of the different call types present in Set B.}
    \label{tab:setBdistribution}
    \begin{tabular}{|ccccccc|}
    \hline
         \textbf{agg}&\textbf{cc}  & \textbf{al} & \textbf{ld} & \textbf{sn} & \textbf{soc} & \textbf{mo}\\
         \hline
         375& 1477 & 645 & 164 & 1854  & 1154  & 759\\
         \hline
    \end{tabular}
  
\end{table}

\subsection{Experimental set-up} \label{subsec:experimental-setup}

\begin{figure}[t]
\centering
  \includegraphics[width=\linewidth]{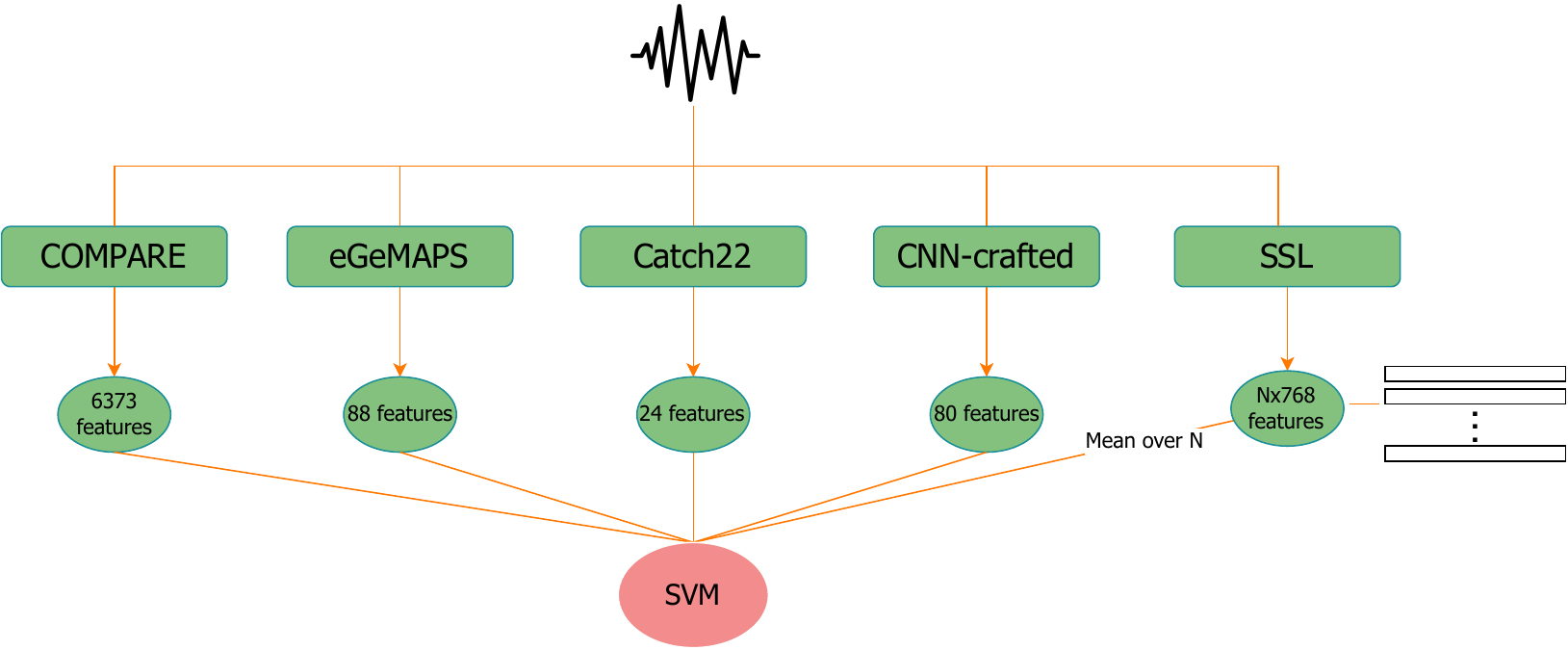}
  \caption{Diagram of the workflow of the study. N denotes number of frames.}
  \label{fig:diagram}
\end{figure}

As a preprocessig step, we downsampled all waveforms to \SI{16}{\kilo\hertz} and vocalizations shorter than \SI{100}{\milli\second} were systematically replicated until they reached the desired minimum duration of 100 ms.  To compare the feature representations, we adopted a 5-fold cross-validation strategy by employing 80:20 train-test split. Figure~\ref{fig:diagram} shows the call classification framework. As illustrated in the figure, a call-level fixed length representation is obtained for each feature type and fed as input to a support vector machine (SVM) based classifier. We compare the feature representations by evaluating the respective call classifiers in terms of unweighted average recall (UAR). We chose UAR as metric due to class imbalance in the datasets.
Unlike weighted average accuracy (classification accuracy), UAR measure gives importance to recognition of all classes. Higher UAR means higher recall across classes.
When training the SVM classifier, we applied a grid search methodology on the training set of each fold with the Unweighted Average Recall (UAR) as the optimization criterion to search space of the hyperparameters (presented in~Table~\ref{tab:svm_hyperparameters}). In the reminder of the section, we explain the call-level fixed length representation obtained for each feature representation type.
\begin{table}[!htb]
\centering
\caption{SVM hyperparameters grid}
\begin{tabular}{|c|c|}
\hline
\textbf{Parameter} & \textbf{Values} \\ \hline
C & 1e[-1, 0, 1, 2] \\ \hline
Gamma  & 1e[-3,-2,-1,0]\\ \hline
Kernel & [`Linear', `RBF', `Polynomial', `Sigmoid'] \\ \hline
\end{tabular}
\label{tab:svm_hyperparameters}
\end{table}

In the case of knowledge-based feature representation, (a) \textit{pycatch22} toolkit was employed for extracting 24 dimensional call-level Catch22 features and (b) openSMILE~\cite{opensmile} tool is used to extract 6373 dimensional call-level COMPARE feature representation and to extract 88 dimensional call-level eGeMAPS feature representations.

In the case of SSL feature representations, the call-level 768 dimensional feature representation is obtained as follows: (a) 768 dimension output of CNN encoder, $1^{st}$, $2^{nd}$, $6^{th}$ or the last transformer layer is obtained per frame and averaged over frames, (b) the 768 dimension output of each of the 12 transformer layers are averaged per frame and then the resulting per frame representation is averaged over frames. The S3PRL toolkit~\cite{sslembeddings} was used to extract the embeddings.

In the case of CNN-crafted feature representation, there is a need to train a CNN-based call classifier for feature extraction. As the data sets were small in size with severe class imbalance, as opposed to training a CNN feature extractor per fold, we employed stratified k-folds cross-validation strategy to get a single CNN feature extractor. This method constructs folds while maintaining class proportion integrity, i.e., ensuring consistent class proportions in both training and test sets, mirroring those of the original dataset. We set the number of folds to 5 and trained CNNs for each fold using the architecture presented in Table~\ref{tab:cnn_architecture} using PyTorch. The adaptive average layer target size was set to one. This allows the network to handle variable length waveform inputs and yield fixed-length (80-dimensional) call level feature representation. We employed the cross-entropy error criterion to train the CNN.
The CNN of the best performing fold was selected to extract 80 dimensional call-level CNN-crafted feature representation (from the output of the fully connected hidden layer).

\begin{table}[!htb]
  \centering
  \caption{CNN architecture for CNN-crafted feature extraction. $n_f$ denotes number of filters. HU denotes number of hidden units.}
  \label{tab:cnn_architecture}
  \setlength{\tabcolsep}{1.5pt} 

  \begin{tabular}{@{}llllll@{}}
    \toprule
    Block & Operation & Kernel & Stride & Padding & $n_f$/HU \\
    \midrule
     & Convolution & 40 & 30 & 0 & 40 \\
       1 & Max Pooling & 2 & 2 & 0 & -\\
      & ReLU Activation & - & - & -&- \\
    \addlinespace
     & Convolution & 7 & 1 & 0 & 40\\
    2  & Max Pooling & 2 & 2 & 0 &- \\
      & ReLU Activation & - & - & -&- \\
    \addlinespace
     & Convolution & 3 & 1 & 0 &80\\
    3  & Max Pooling & 2 & 2 & 0 &-\\
      & ReLU Activation & - & - & -&- \\
    \addlinespace
     & Adaptive Avg Pooling & - & - & -&- \\
    4  & Flatten & - & - & - &-\\
      & Fully Connected & - & - & - &80\\
    \bottomrule
  \end{tabular}
\end{table}

\begin{figure*}[!htb]
\centering

  \includegraphics[trim={5 5 5 5},width=0.24\linewidth]{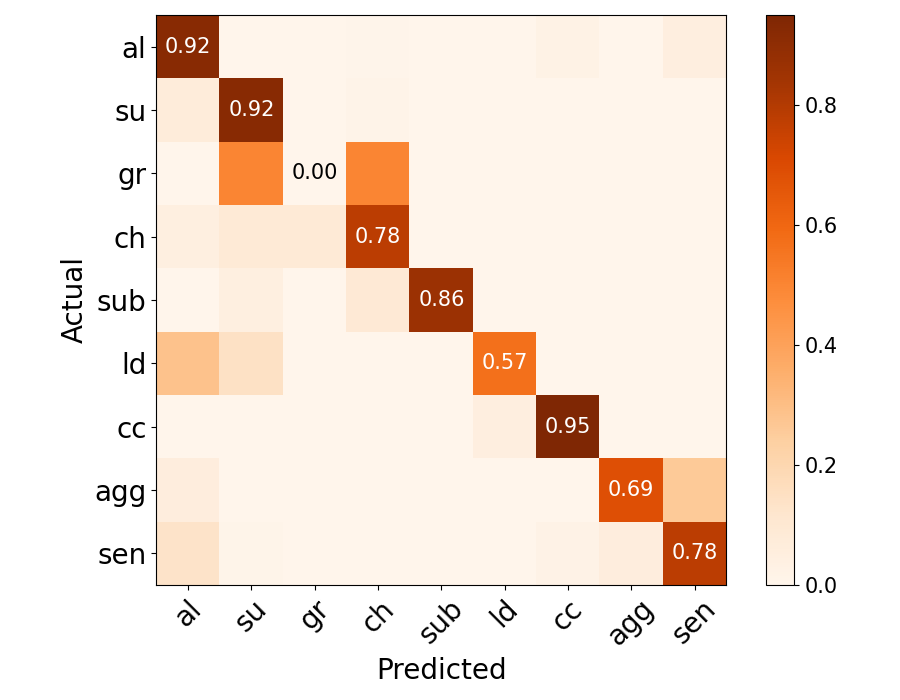}
  \includegraphics[trim={5 5 5 5},width=0.24\linewidth]{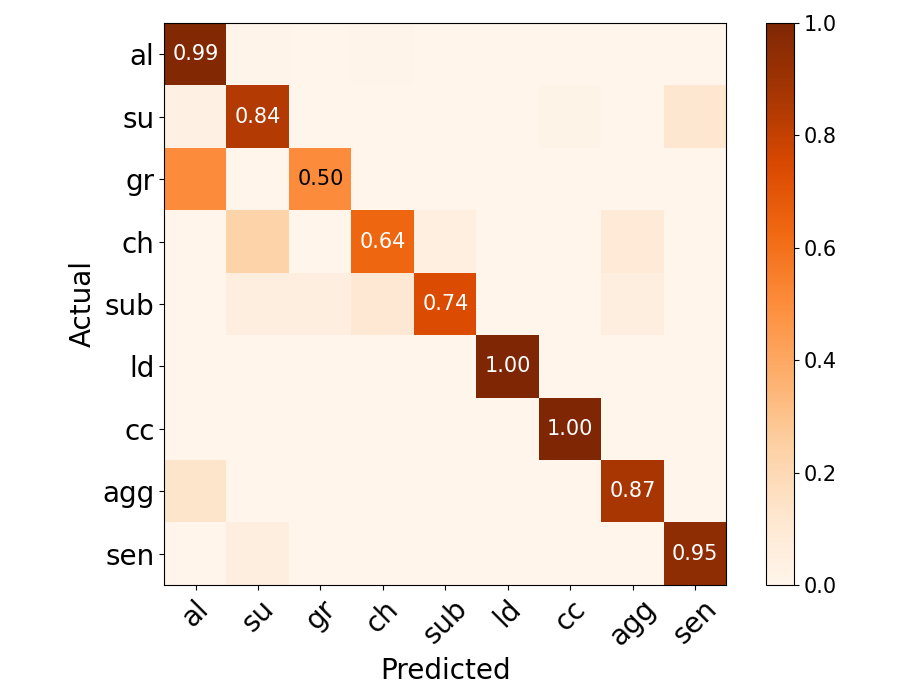}
  \includegraphics[trim={5 5 5 5},width=0.24\linewidth]{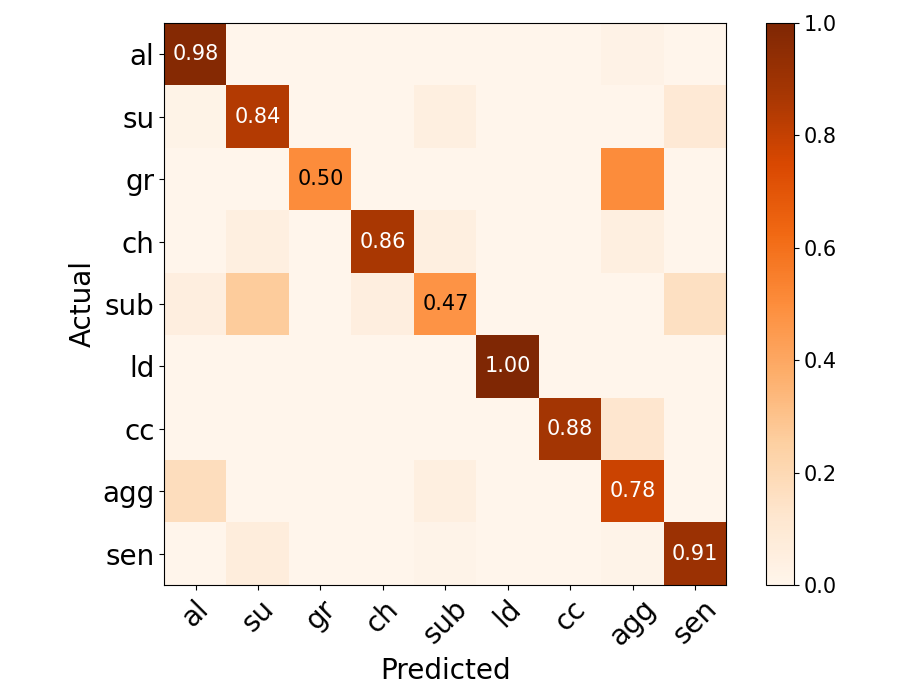}
  \includegraphics[trim={5 5 5 5},width=0.24\linewidth]{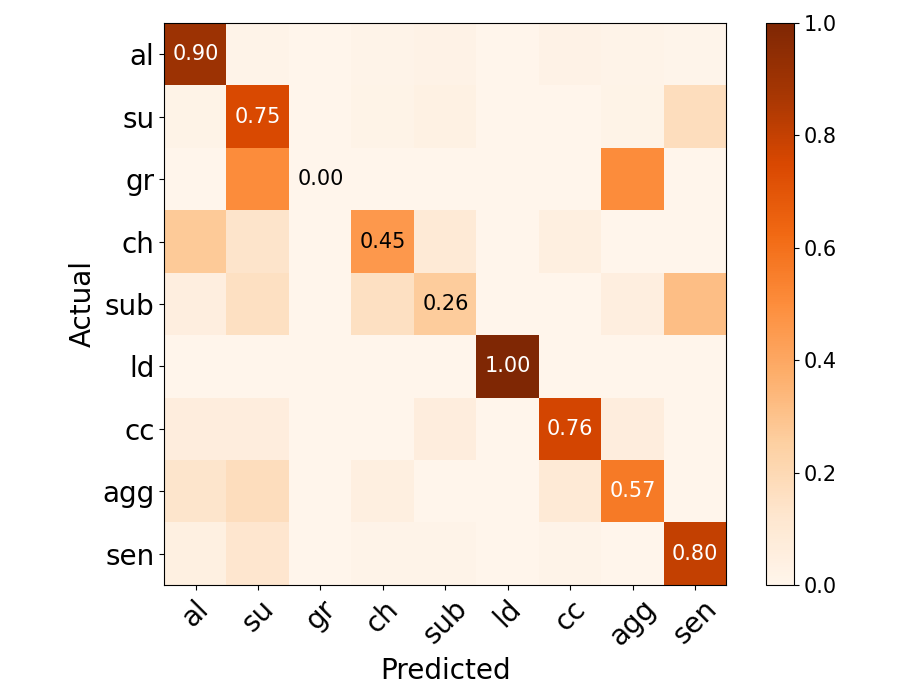}
  \caption{Confusion matrices for SVM classifier using, \textit{from left to right}, WavLM, CNN-crafted, COMPARE and Catch22 embeddings on the test set of Set A.}
  \label{fig:conf_matrix_setA}
\end{figure*}

\begin{figure*}[!htb]
\centering

  \includegraphics[width=0.24\linewidth]{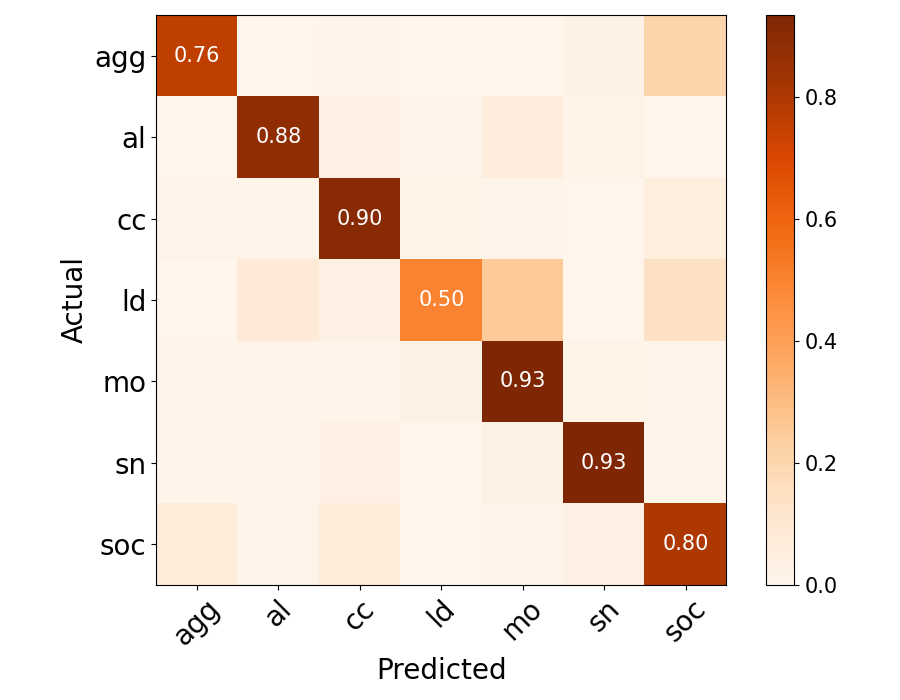}
  \includegraphics[width=0.24\linewidth]{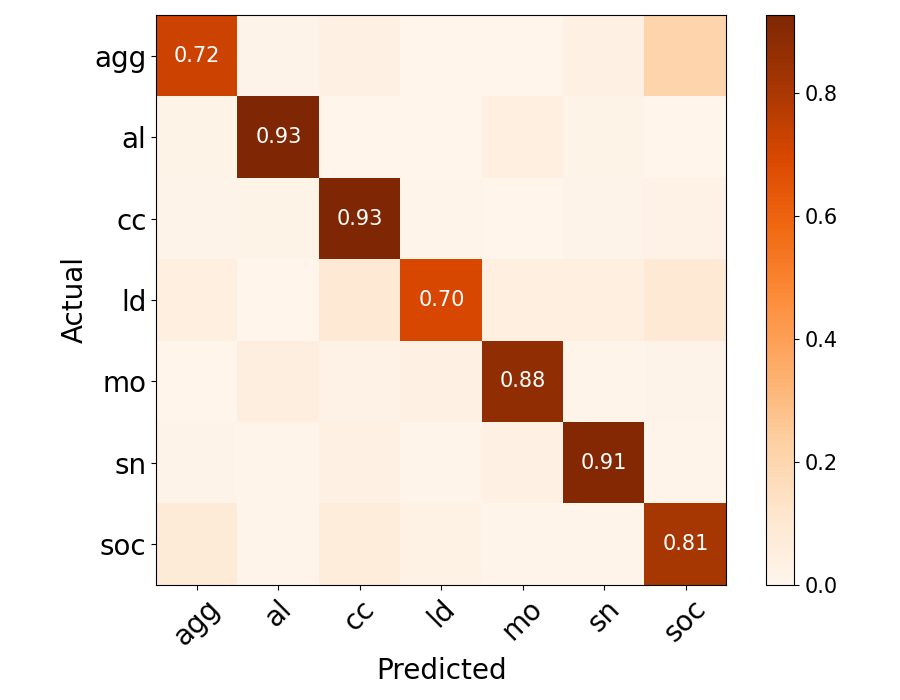}
  \includegraphics[width=0.24\linewidth]{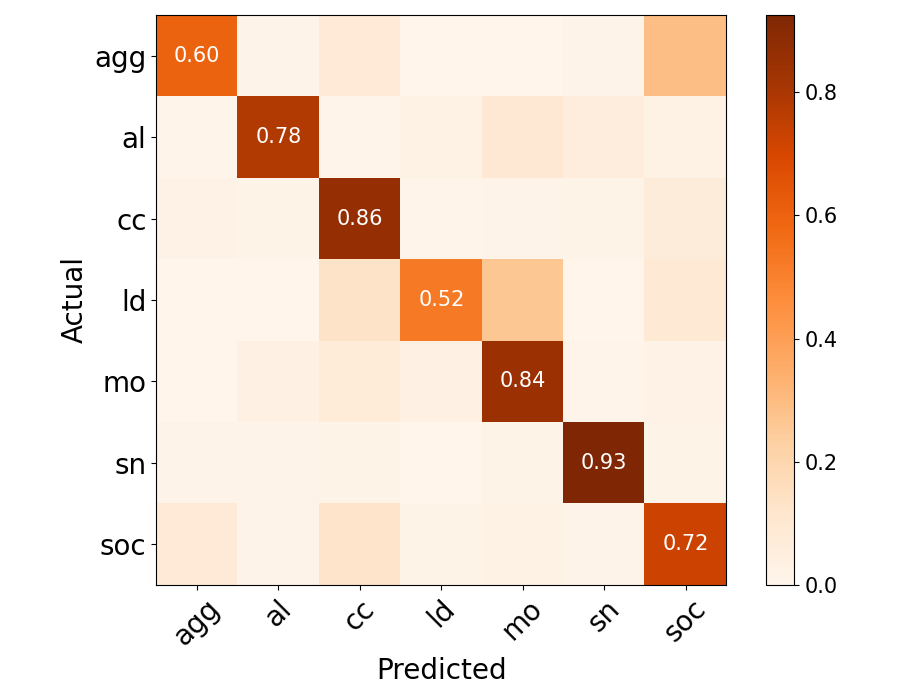}
  \includegraphics[width=0.24\linewidth]{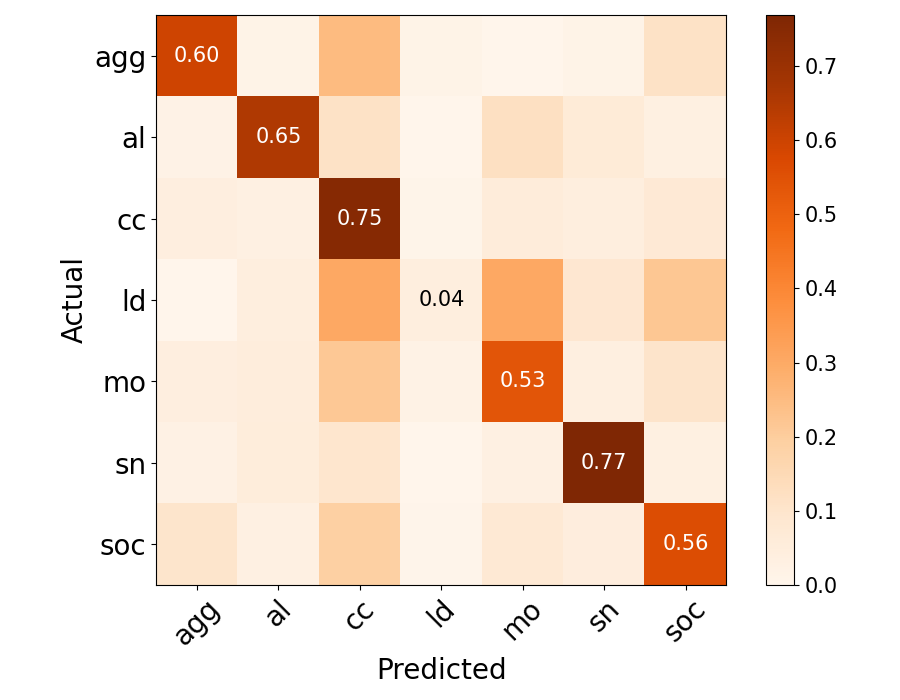}
  \caption{Confusion matrices for SVM classifier using, \textit{from left to right}, WavLM, CNN-crafted, COMPARE and Catch22 embeddings on the test set of Set B.}
  \label{fig:conf_matrix_setB}
\end{figure*}

\section{Results and discussion}\label{sec:results-discussion}

Table~\ref{tab:combined_results} presents an analysis of SSL neural embeddings. It can be observed the that lower layer transformer layer embeddings and CNN encoder representations yield better systems than higher layer transformer layer embeddings. Averaging the embeddings across the transformer layers, although yields better system than layer $6$ and last layer embeddings, is not helpful when compared to layer $1$ embedding, layer $2$ embedding or CNN encoder output alone. Taken together, this indicates that lower transformer layer embeddings of SSLs pre-trained on human speech are more informative than higher transformer layer embeddings for meerkat call classification.


\begin{table}[!htb]
\centering
\caption{UAR scores of chosen representations using wav2vec2 (W2), WavLM (WL) and HuBERT (HT) models on Test set of Set A and B}
\label{tab:combined_results}
\resizebox{\columnwidth}{!}{%
\begin{tabular}{c|ccc|ccc}
\hline
\multicolumn{1}{c}{} & \multicolumn{3}{c}{Set A} & \multicolumn{3}{c}{Set B} \\
\cline{2-7}
Model & W2 & WL & HT & W2 & WL & HT \\
\hline
CNN & $0.71$ & $0.68$ & $0.74$ & $0.78$ & $0.77$ & $0.78$ \\
$1^{st}$ Transformer & $0.71$ & $0.72$ & $0.73$& $0.79$ & $0.82$ & $0.78$ \\
$2^{nd}$ Transformer & $0.73$ & $0.71$ & $0.72$ & $0.79$ & $0.82$ & $0.79$ \\
$6^{th}$ Transformer & $0.54$ & $0.50$ & $0.64$& $0.69$ & $0.70$ & $0.76$\\
Last Transformer & $0.35$ & $0.38$ & $0.55 $ & $0.52$ & $0.53$ & $0.67$ \\
Average of Transformers & $0.63$ & $0.59$ & $0.61$ & $0.75$ & $0.72$ & $0.76$ \\
\hline
\end{tabular}%
}
\end{table}


Table~\ref{tab:result} compares the systems across different feature representations. For SSL feature representation wav2vec2, WavLM and HuBERT, we have reported the best system performance from Table~\ref{tab:combined_results}. In the case of hand-crafted features, it is observed that eGeMAPS and COMPARE feature based systems yield better system than Catch22 feature representation. In the case of SSL feature representations, the systems are comparable. The CNN-crafted feature representation yields the best systems. When comparing hand-crafted features and neural embeddings, COMPARE feature outperforms SSL features on Set A and performs slightly worse when compared to wav2vec2 and HuBERT. It is worth pointing out that the COMPARE feature largely outperforms higher transformer layer embedding based systems (layer $6$ and last layer in~Table~\ref{tab:result}). This indicates that, similar to neural embeddings from networks pre-trained on human speech, hand-crafted representations developed for speech processing applications can be useful for meerkat call classification. 

\begin{table}[!htb]
\centering
 \caption{UAR scores on Test set of Set A and B with 5-fold CV for call types classification}
    \label{tab:result}
    \begin{tabular}{ccc}
    \hline
    \vspace{0.1cm}
      Model & Set A & Set B \\
         \hline
         eGeMAPS & 0.61 & 0.66 \\
         COMPARE & 0.80& 0.75 \\
         Catch22 & 0.61& 0.56\\
         
         \hline
         
         wav2vec2 & 0.73 & 0.79\\
         WavLM & 0.72 & 0.82\\
         HuBERT & 0.74 & 0.79 \\ 
         CNN-crafted & \textbf{0.84} & \textbf{0.84}\\
         
         \hline
    \end{tabular}
\end{table}
The main distinction between Set A and Set B lies in the number of classes, the number of samples, and the class distribution within the datasets. As discussed previously, Set B comprises more samples, fewer number of classes and exhibits better class balance than Set A. Therefore, our initial expectation was that Set B would yield superior performance. This hypothesis is confirmed with the SSL models, eGeMAPS, and the CNN model, where results with Set B perform better than Set A. Confusion matrices for WavLM, CNN-crafted, COMPARE and Catch22 are presented for Set A and Set B in Figure~\ref{fig:conf_matrix_setA} and Figure~\ref{fig:conf_matrix_setB}. It can be observed that all the call types are mostly classified well except for "gr" in Set A which has the lowest amount of data.

\begin{figure}[!htb]
 \centering
 \includegraphics[width=0.98\linewidth]{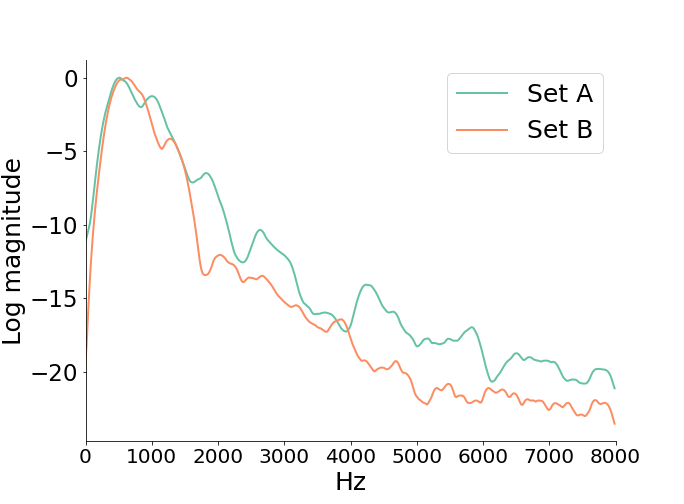}
 \caption{Cumulative frequency responses of first layer filters of CNN}
 \label{fig:kernelwidth}
 \end{figure}

For the case of CNN-crafted, Figure~\ref{fig:kernelwidth} shows the cumulative frequency response of the 40 first layer convolution filters. This is estimated by applying a DFT of 1024 points on filters of length 40 samples and taking logarithm of the summed magnitude responses. Although Set A and Set B have been collected independently and labeled, it can be observed that the cumulative filter responses of the CNNs of Set A and Set B are similar with a major emphasis between 0-\SI{2}{\kilo\hertz}. This indicates that the CNNs are capturing information systematically for class classification across the two data sets. In our future work, we will investigate what kind of acoustic information does that frequency range carries in meerkat vocalizations for call analysis.

\section{Conclusions} \label{sec-conclusions}
Meerkats with their highly social nature and diverse vocal repertoire, provide an intriguing model system for investigating animal communication and, as an extension could help us better understand the evolution of human communication. One of the challenges in that direction is the lack of methods for automatic meerkat call analysis. In that direction, this paper explored feature representations for automatic analysis of meerkat vocalizations. We compared time-series analysis-based hand-crafted feature representation, hand-crafted feature representations developed for human speech processing, SSL-based feature representations obtained from neural networks trained on human speech, and feature representations automatically learned in a task-dependent manner from meerkat calls using CNNs. Our studies show that hand-crafted feature extractors and SSL feature extractors developed for human speech processing can be used for meerkat call classification. Similarly, we observe that the CNN-based method developed for automatic feature learning in a task-dependent manner for human speech processing can be scaled for meerkat call classification task (CNN-crafted). Our future work will focus on analyzing these diverse feature representations to tease out and explain the acoustic information that is relevant for meerkat call analysis.

\section{Acknowledgement}
The first author carried out a part of this work at Idiap as part of her Master-AI thesis at UniDistance, Switzerland. This work was partially funded by Swiss National Science Foundation’s NCCR Evolving Language project (grant no. 51NF40 180888).

\clearpage

\bibliographystyle{IEEEtran}
\bibliography{mybib}

\end{document}